\documentclass[12pt]{article}
\usepackage{graphicx}
\def\pa{\alpha^\prime}

\def\th{\theta}
\def\S{\Sigma}
\def\s{\sigma}
\def\f{\phi}

\def\G{\Gamma}

\def\d{\delta}

\def\l{\lambda}
\def\L{\Lambda}

\def\th{\theta}

\def\i{\iota}

\def\pa{\partial}

\def\cd{{\cal D}}

\def\ha{\frac12}

\newcommand{\be}{\begin{equation}}
\newcommand{\ee}{\end{equation}}
\newcommand{\bea}{\begin{eqnarray}}
\newcommand{\eea}{\end{eqnarray}}

\def\dim{\textrm{dim}}
\begin{document}

\bigskip
\begin{center}
{\bf{\Large Errata to ``Flat Spacetime Vacuum in Loop Quantum Gravity" }}
\end{center}

\bigskip
\bigskip
\centerline{A. MIKOVI\'C}
\centerline{Departamento de Matem\'atica, Universidade Lus\'ofona}
\centerline{Av. do Campo Grande, 376, 1749-024, Lisboa, Portugal}

\bigskip
\bigskip
\begin{abstract}
{\small We give the correct expressions for the spin network evaluations proposed in \textit{Class. Quant. Grav. {\bf 21} (2004) 3909} as the coefficients of the quantum gravity vacuum wavefunction in the spin network basis.}
\end{abstract}

\bigskip
\bigskip
\bigskip
In \cite{vlqg} a proposal for the construction of the flat spacetime vacuum wavefunction in the loop quantum gravity formalism was given. According to this proposal, if the spacetime manifold has a topology $\S\times \bf R$ and $A$ is a flat complex connection one-form
on the $SU(2)$ principal bundle over $\S$, then a holomorphic gauge-invariant functional $\Psi[A]$ satisfies the Ashtekar constraints and hence it can be considered as a physical wavefunction. In order to obtain a physical wavefunction which describes the flat background geometry, one should chose $\Psi[A]$ such that it is peaked around the flat background values of the triads. The simplest choice is 
\be\Psi[A]=\exp\left(i\,Tr\int_\S E^k_0 (x) A_k (x) d^3 x \right)\,,\ee 
where the components of $E_0(x)$ represent the background triads. Although this choice is not gauge invariant, it can be modified when the theory is discretized such that one obtains a gauge invariant result. The corresponding state in the spin network basis is given by
\be |\Psi\rangle = \sum_s I(s)\,|s\rangle \,, \ee
where $|s\rangle$ is an orthonormal basis of states labeled by the $SU(2)$ spin networks $s$,
and
\be I(s)=\int \cd A \, W_s [A]\,\d(F)\,\Psi[A] \,,\ee
where $W_s [A]$ is the spin network wavefunction and $F$ is the connection curvature.

The path-integral $I(s)$ can be defined by discretizing $\S$ by using a corresponding triangulation. This gives
\be I(s)=\int \prod_l dg_l \,W_s (g_{l'})\,\prod_f \d (g_f)\,\prod_l e^{iTr(A_l E_l^0)} \,,\ee
where $l$ and $f$ denote the edges and the faces respectively of the dual two-complex $\G$, $g_l$ denote the corresponding $SU(2)$ holonomies, $g_f=\prod_{l\in\pa f}g_l$ and $l'$ denote the dual edges which carry the irreps $j_s$ of the embedded spin network $s$. After performing the group integrations, one obtains the following state-sum
\be I(s)=\sum_{\L_f,\l_l,\i_l}\prod_f \dim\,\L_f \,\langle \,\prod_l C(E_l^0,\l_l)\prod_v A_v (\L_f,j_s,\i_s,\i_l)\,\rangle \,,\label{is}\ee
where $C$ are the edge amplitudes and $A_v$ are the vertex amplitudes of the spin foam $\G$. 
The $\i_s$ and $\i_l$ are the corresponding intertwiners, while the expression
$$ \langle \,\prod_l C(E_l^0,\l_l)\prod_v A_v (\L_f,j_s,\i_s,\i_l)\,\rangle$$
indicates the evaluation of the spin network obtained by composing $\G$ and $s$. The corresponding graph is obtained by inserting the tetrahedron graphs into the vertices of the graph of $\G$ and then by inserting the vertices of the spin network $s$ into the corresponding tetrahedra, see Fig. 1. All the tetrahedra carry the irreps $\L_f$ and the tetrahedra which contain the  vertices of the spin network $s$ have the additional edges carrying the corresponding irreps $j_s$. Each edge $l$ of $\G$ carries an irrep $\l_l$ and has the insertion $C(E_l^0,\l_l)$. 

In \cite{vlqg}, the formula (\ref{is}) was presented, but it was not as precisely defined as it is done here. Consequently, the expression for $I(s)$ in the case of the theta-5 spin network embedded in the triangulation of $\S=S^3$ consisting of two tetrahedra was not correct, because it was based on the wrong graph. The correct expression is based on the evaluation of the spin network whose graph is given in Fig. 1.
\begin{figure}
\centering
\includegraphics{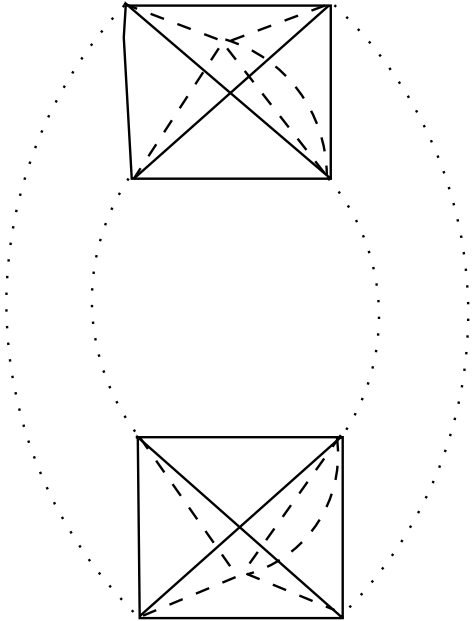}
\caption{\small{The graph corresponding to a theta-5 spin network embedded into the  triangulation of $S^3$ consisting of two tetrahedra. The solid lines carry the face irreps $\L_f$, the dashed lines carry the embedded spin network irreps $j_s$ and the dotted lines carry the edge irreps $\l_l$.}}
\end{figure}

The second error appeared in the evaluation of the $C(E_l^0,\l_l)$ matrix elements. The average value of an  element of the matrix $C(E,j)$ was defined as 
\be C_j (E) = {Tr\,C(E,j)\over\dim\,j} \,,\ee 
which was determined by the group integral
\be C_j (E) = \int_{SU(2)} dg \,\chi^{(j)}(g)\,e^{i\vec A\cdot\vec E} \,,\quad g=e^{i\vec A\cdot\vec\s}\,,\label{cj}\ee
where $\chi^{(j)}$ is the representation character and $\vec\s$ are the Pauli matrices. However, when calculating (\ref{cj}), an incorrect expression for the $SU(2)$ group measure was used. 
The correct expression is given by
\be dg = {1\over 8\pi^2} \,\sin^2(A/2)\,\sin\th \,dA\, d\th\,d\f\,,\ee
where $A=|\vec A |\in(0,4\pi)$, $\th\in(0,\pi)$ and $\f\in(0,2\pi)$, so that
\be C_j (E) = {1\over 4\pi}\int_0^{4\pi} dA\int_0^\pi \sin\th\,d\th\,\sin^2(A/2) \,{\sin(j+\ha)A\over\sin(A/2)}\,e^{iEA\cos\th}\,,\ee
where $E=|\vec E |$. This gives
\be C_j (E) = {1\over 2\pi}\int_0^{4\pi}dA\,\sin(A/2) \,\sin\left[\left(j+\ha\right)A\right]\,{\sin (EA)\over EA}\,,\ee
which can be expressed as
\be C_j (E) = {1\over 8\pi E}\left[K(j+E)-K(j-E)-K(j+1+E)+K(j+1-E)\right]\,,\ee
where
\be K(a)=\int_0^{4\pi a}\,{\sin x\over x}\,dx \,.\ee

\end{document}